\documentclass[%
 nofootinbib,
 reprint,
 amsmath,amssymb,
 aps,
]{revtex4-2}

\pdfoutput=1 

\usepackage{graphicx}
\usepackage{dcolumn}
\usepackage{bm}

\usepackage{algorithm} 
\usepackage{algpseudocode}  

\begin{document}


\title{Adapted Caldeira-Leggett Model}

\author{Andreas Albrecht}
 \email{ajalbrecht@ucdavis.edu}
 \author{Rose Baunach}
 \email{baunach@ucdavis.edu}
\affiliation{Center for Quantum Mathematics and Physics and Department of Physics and Astronomy\\ UC Davis, One Shields Ave, Davis CA.}

\author{Andrew Arrasmith}%
 \email{aarrasmith@lanl.gov}
\affiliation{Theoretical Division, Los Alamos National Laboratory, Los Alamos, NM USA.}


\date{\today}

\begin{abstract}
We preset a variant of the Caldeira-Leggett (CL) model of a harmonic oscillator coupled to an environment. The CL model is a standard tool for studying the physics of decoherence. Our ``adapted Caldeira-Leggett'' (ACL) model is built in a finite Hilbert space which makes it suitable for numerical studies.  Taking a numerical approach allows us to avoid the limitations of standard approximation schemes used with the CL model.  We are able to evolve the ACL model in a fully reversible unitary manner, without the built-in time asymmetry and other assumptions that come with the master equation methods typically used.  We have used the ACL model to study new topics in the field of decoherence and einselection where the full unitary evolution is essential to our work.  Those results (reported in companion papers) include an examination of the relationship between einselection and the arrow of time, and studies of the very earliest stages of einselection. This paper provides details about the ACL model and our numerical methods.   Our numerical approach makes it straightforward to explore and plot any property of the physical system. Thus we believe the examples and illustrations we present here may provide a helpful resource for those wishing to improve their familiarity with standard decoherence results, as well as those looking to probe the underpinnings of our companion papers.  We expect the ACL model will be a useful tool for exploring additional phenomena that cannot be studied using traditional approximation schemes. 
\end{abstract}

\maketitle


\section{\label{sec:Intro}Introduction}

The Caldeira-Leggett (CL) model is a toy model describing a particle which moves in its own potential and is also coupled to an environment~\cite{Caldeira:1981rx,Caldeira:1982iu,Caldeira:1982uj}. The environment is usually treated as an infinite set of harmonic oscillators, and the particle is often taken to move in a harmonic potential as well. The particle plus environment describe a closed system which can in principle be treated quantum mechanically as a system undergoing reversible unitary evolution. In practice the CL model is often treated in the ``Markovian limit'' where the particle evolution can be described by an irreversible master equation. Working in this limit provides tractable mathematics which can be used to study particle-environment interactions in situations which naturally have an arrow of time.  For example the CL model has been used in pioneering explorations of  decoherence~\cite{Zurek1986} and einselection~\cite{Zurek:1992mv}. 

This paper introduces an ``adapted Caldeira-Leggett'' (ACL) model.  The adaptations are chosen to reproduce the essential features of the CL model as fully as possible within a finite Hilbert space.  The goal is to be able to evolve the ACL model easily on a desktop computer in its full unitary form, thus enabling the convenient exploration of a more complete range of physical situations including those outside the Markovian limit.   

Aside from describing various technicalities of how we construct the ACL model, we present here results from ``putting it through its paces'' which demonstrate that the ACL model does a good job of reproducing physics phenomena that are an established part of the decoherence literature. These cross-checks give us a solid foundation on which to explore the new directions, which we report in companion papers~\cite{ACLeqm,CopyCat}. For the most part, we do not expect the phenomena presented in this paper to be new to an expert on decoherence.  On the other hand, someone learning this topic might find our graphical presentation centered on a specific physical system a useful compliment to a more thorough review such as~\cite{schlosshauer2007decoherence} and may even provide a helpful starting point.  

The physics of einselection plays an important role in many physical phenomena (see for example~\cite{schlosshauer2007decoherence,Zurek:2022fah,Strasberg:2023gdl}). The development of the ACL model was originally motivated by our interest in exploring the physics of einselection under equilibrium conditions\footnote{These motivations originate in cosmology where connections between the emergence of classicality (related to einselection) and the arrow of time (which originates with cosmology, as discussed for example in~\cite{Albrecht:2014eaa}) might lead to useful insights.}.  The Markovian limit, with its definite arrow of time, clearly cannot describe the full fluctuations of an equilibrium system.
We also expect the ACL model will be useful in exploring other physics outside of the Markovian regime, and we have already found one such example (which we've named the ``copycat process'') that we mention briefly in Sect.~\ref{sec:Copycat} and develop further in a companion paper~\cite{CopyCat}. We have also found the ACL model useful for exploring notions of thermalization in finite systems~\cite{Albrecht:2022mrx}.

While there are a variety of other methods that can also go beyond the limitations of Markovian evolution (see e.g. this review~\cite{deVega2017} and references therein), our goal was to specifically model einselection in a clear and transparent manner with as few computational resources as possible. We found that a simplistic model of the environment (as a general scrambler following~\cite{Albrecht:1992rs,Albrecht:1993hf}) helped realize these priorities (versus basing the environment on a detailed physical system~\footnote{Still, we note some connections between the ACL model and NMR systems in Sects.~\ref{sec:RCL} and~\ref{sec:NMR_Loschmidt} and also in~\cite{CopyCat}.}). Also, since the CL model is one of the pioneering models of einselection, it made sense to develop the ACL model, to better compare with the existing decoherence literature. 

We organize this paper as follows. Section~\ref{sec:Basic} defines the ACL model and demonstrates the robustness of our numerical calculations. Section~\ref{sec:Examples} explores a variety of standard results from the literature using the ACL model.  For example we show how an initial Schr\"odinger cat state of superposed wavepackets is einselected to a classical mixture of single packets. We also introduce the ``copycat process,'' a new phenomenon which we explore extensively in~\cite{CopyCat}. Section~\ref{sec:Approach2Eqm} explores the way the ACL model both approaches and then remains solidly situated in a fluctuating equilibrium state when evolved long enough.  The presence of a fully fluctuating equilibrium state is a behavior not accessible through master equation techniques, but one which is very naturally achieved with our methods.  This equilibrium behavior forms a foundation for our exploration of the relationship between einselection and the arrow of time in~\cite{ACLeqm}. In Sect.~\ref{sec:RCL} we introduce the ``reduced Caldeira-Leggett'' (RCL) model which replaces the SHO with a single qubit. We demonstrate how the RCL model can access a different set of phenomena.  The results from this paper are placed in the context of the existing literature in Sect.~\ref{sec:Comparison}. Among other things, we relate some of our results to Nuclear Magnetic Resonance (NMR) physics and ``Loschmidt echos'' (a concept developed in discussions of the arrow of time).  Section~\ref{sec:Conclusions} presents our conclusions.  

A series of appendices present additional technical information.  Appendix~\ref{sec:QuntLim} explores einselection in the ``quantum limit'' of the ACL model. Appendix~\ref{sec:EigHs} presents details of the eigenstates of the truncated SHO, which reveal differences between the truncated and the continuum cases. We give a detailed picture of the spectra of the different Hamiltonians (SHO, environment and combined) in Appendix~\ref{sec:spectra}.  Appendix~\ref{sec:Numerical} presents our numerical techniques and tolerances.

\section{\label{sec:Basic}The ACL model}
\begin{figure}
    \centering
    \includegraphics{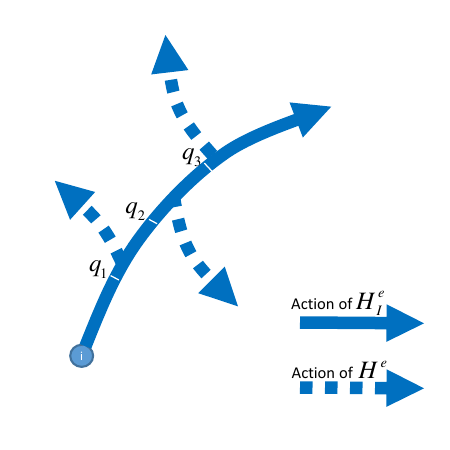}
    \caption{The interaction term $q\otimes H_e^I$ moves the initial environment state along a specific path in the $e$ Hilbert space determined by $H_e^I$, illustrated by the solid curve. The rate of movement along this path is proportional to the value of $q$, and that allows different $q$ states to become entangled with different environment states.  In the case where $\left[ H_e^I, H_e \right] \neq 0$, the action of $H_e$ can push the evolution off the original path in a variety of different directions depending on the starting point ($\propto$ the value of $q$). These various paths are illustrated by the dashed curves.  The non-commuting property can make the process of entanglement much more efficient (especially for the large $N_e$ case, not shown in this sketch).  }
    \label{fig:HeHI}
\end{figure}

We consider a ``world'' Hilbert space $w = s\otimes e$ which is a tensor product of a ``system'' Hilbert space $s$ and the environment space $e$.  We consider a Hamiltonian of the form
\begin{equation}
H_w = H_s \otimes {\bf{1}}_{}^e + {q_s}\otimes H_e^I + {{\bf{1}}^s} \otimes {H_e}.
\label{eqn:Hform}
\end{equation}
Equation~\ref{eqn:Hform} describes the form of both the CL and ACL models. The differences arise in the specifics of the different ingredients. These are the system Hamiltonian $H_s$, the self-Hamiltonian for the environment $H_e$,  and the piece of the interaction Hamiltonian in the $e$ subspace, $H_e^I$. We focus on the case where $s$ is a simple harmonic oscillator (SHO). The position operator of the system, $q_s = q_{SHO}$, is defined in the usual way for the CL model. However for the ACL model $s$ is a ``truncated SHO'' (in order to allow a numerical treatment) and the definition of $q_{SHO}$ for that case is nontrivial. 
Hamiltonians of this form have features that enable the system to become entangled with the environment in ways that reflect certain realistic physical situations. The interaction term changes the state of the environment with a strength proportional to the value of $q_{s}$, so different positions become entangled with different environment states.  When $H_e$ and $H_e^I$ don't commute (the case for both CL and ACL models) the entangling process is much more effective, as illustrated heuristically in Fig.~\ref{fig:HeHI}. 
\subsection{\label{sec:selfHs} The SHO}
\begin{figure}
    \centering
    \includegraphics{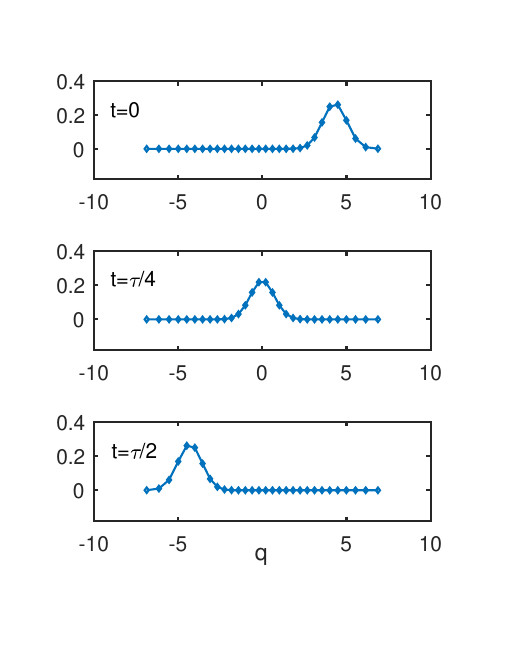}
    \caption{A coherent state wavefunction (squared) for the truncated SHO shown at different points in its period $\tau$.  Despite certain differences from the continuum case noted in the text, the shape and robustness under evolution of this state corresponds to the properties of continuum coherent states.}
    \label{fig:SHOcstestB}
\end{figure}

\begin{figure}
    \centering
    \includegraphics{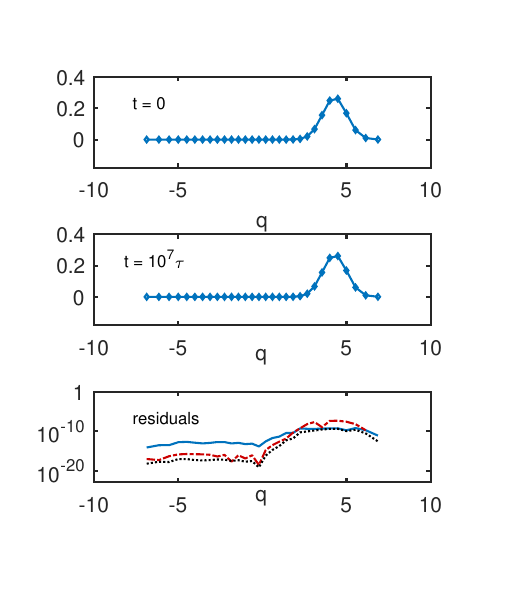}
    \caption{A coherent state at $t=0$ and $t = 10^7\tau$.  The third panel shows the residuals for the probabilities (solid) and for the real (dotted) and imaginary (dot-dashed) part of the amplitude.  These curves illustrate that the numerically evolved truncated model reproduces the periodic properties expected of the continuum case to an excellent degree of accuracy.}
    \label{fig:SHOcstest}
\end{figure}

For a normal (un-truncated) SHO the matrix elements of the lowering operator $\bf{a}$ in the basis given by number (or energy) eigenstates is given by 
\begin{equation}
    \left\langle i \right|{\bf{a}}\left| j \right\rangle  = \sqrt j {\delta _{i,j - 1}}
\end{equation}
with $j\geq1$.  For our truncated SHO the same formula is valid for ${\bf{\hat a}}$ (where the hat denotes the truncated version) but it only applies for $\left\{ {i,j \in 1:{N_s}} \right\}$ where $N_s$ is the size of the truncated SHO Hilbert space. The operator $\bf{{\hat a}^\dagger}$ is formed by conjugating ${\bf{\hat a}},$ and ${\bf{\hat q}}$, ${\bf{\hat p}}$  and ${\bf{\hat H_{SHO}}}$ are all constructed from ${\bf{\hat a}}$ and $\bf{{\hat a}^\dagger}$ using the usual formulas from the un-truncated case. These operators in the truncated space don't have all the usual properties due to the truncation. For example 
\begin{equation}
    \left[ {{{\bf{\hat a}} },{\bf{\hat a}}^\dag} \right] = {\bf{1}} + \Delta
\end{equation}
where $\Delta(i,j) = -N_s\delta_{i,N_s}\delta_{j,N_s}$.  We chose these definitions for the truncated operators because they have some practical advantages over other choices.  The main advantage is illustrated in Figs.~\ref{fig:SHOcstestB} and~\ref{fig:SHOcstest}.  
Figure~\ref{fig:SHOcstestB} shows a coherent state constructed thus:
\begin{equation}
\psi_\alpha \left( q \right) = \left\langle {q}
 \mathrel{\left | {\vphantom {q \alpha }}
 \right. \kern-\nulldelimiterspace}
 {\alpha } \right\rangle  = \left\langle q \right|\exp \left( {\alpha {{\bf{\hat a}}^\dag } - {\alpha ^*}{\bf{\hat a}}} \right)\left| 0 \right\rangle 
 \label{eqn:CS}
\end{equation}
where $\left| 0 \right\rangle$ is the ground state of ${\bf{\hat H_{SHO}}}$ and $ \left\langle q \right|$ is the  $q$  eigenstate of  ${\bf{\hat q}}$.  The $x$ axis gives the eigenvalue of ${\bf{\hat q}}$, which is really a discrete quantity (${\bf{\hat q}}$ has only $N_s$ eigenvalues, which run from $-2\pi$ to $2\pi$). The discrete sets of points plotted (shown by markers) are connected only to reference the continuum of the  un-truncated SHO which this system is intended to approximate\footnote{The truncated form does lead to some novel features in the eigenstates of $H_s$ as discussed in Appendix~\ref{sec:EigHs}.}.  We call the SHO period $\tau$ and in our units $\tau = 2\pi$. We've taken $N_s = 30$ here, and in all the examples shown in this paper.  

The top two panels of Fig.~\ref{fig:SHOcstest} show the same coherent state at $t=0$ and $t = 10^7\tau$. The third panel shows the residuals.  The very small sizes of the residuals further demonstrate the robust nature of the truncated SHO.  The specifics of our numerical approach (including several additional checks) are discussed in Appendix~\ref{sec:Numerical}.

\subsection{\label{sec:Hes}The interaction and environment self-Hamiltonian}

The interaction Hamiltonian has the form ${q_{s}}\otimes H_e^I$. For the ACL model we use ${q_{s} = \bf{\hat q}}$. The environment piece, $H_e^I$, has the form 
\begin{equation}
H_e^I = {E_I}R_I^e + E^0_I.
\label{eqn:Hiedef}
\end{equation}
The matrix $R_I^e$ is a random matrix constructed by drawing each of the real and imaginary parts of each independent matrix element of a $N_e\times N_e$ Hermitian matrix uniformly from the interval $[-0.5,0.5]$ using the computer's random number generator.  

The environment self-Hamiltonian is given by 
\begin{equation}
H_e = {E_e}R^e + E^0_e
\label{eqn:Hedef}
\end{equation}
where $R^e$ is constructed in the same manner as $R_I^e$, but as a separate realization. In Eqns.~\ref{eqn:Hiedef} and \ref{eqn:Hedef}, $E_I$ and $E_e$ are c-numbers which parameterize the overall energy scales. Both $R_I^e$ and $R^e$ are fixed initially and are not changed during the time evolution. The full Hamiltonian of the ACL model is time independent. All the results in this paper use $E^0_I=E^0_e=0$, but we have found nonzero values for these offset parameters to be helpful for other calculations we report elsewhere. 

The job of $H_e^I$ and $H_e$ is to move states around in the environment efficiently, so that entanglement between the SHO and the environment can emerge as fully as possible despite working within the confines of a finite system\footnote{The approach to $H_e^I$ and $H_e$ used here is similar to that pioneered in~\cite{Albrecht:1992rs}, although in that work the ``system'' was a single qubit.}.  We find the random form of these operators does this job well, and since $\left[ {H_e^I,{H_e}} \right]$ is just another random matrix the non-commutivity discussed with Fig.~\ref{fig:HeHI} is easily achieved. 
The work presented here uses $N_e=600$. This choice, along with $N_s=30$, was made via an informal optimization process to maximize the utility of the ACL model within the constrained resources of our desktop computer. 

There is also a simple way to modify our ACL model to create $H_e$'s with different spectra.  The crucial aspect achieved by the random matrices in $H_w$ is the non-commutivity of $H_e^I$ and $H_e$.  This aspect is enabled by the eigen\emph{vectors} of independently generated random matrices in large spaces having very little overlap. One could alternatively create $H_e^I$ and $H_e$ by starting in diagonal form (with a spectrum of eigenvalues of your choosing) and then changing basis using a random unitary to produce a ``random matrix'' with the specified eigenvalue spectrum.  We experimented a bit with this approach to generating  $H_e^I$ and $H_e$, but did not find that the extra complexity sufficiently changed the quality of the explorations we were doing to be worthwhile for our purposes.

The next few sections contain some illustrative examples to showcase how standard decoherence phenomena are realized in the ACL model.  We also lay groundwork for new results discussed in more detail in~\cite{CopyCat,ACLeqm}.  The technical minded reader may also wish to refer to Appendices~\ref{sec:EigHs} and~\ref{sec:Numerical}, as these appendices provide more details on the numerical realization of the ACL model (Appendix~\ref{sec:Numerical}), and its sensitivity to the finite dimensional Hilbert space quantities introduced in Sect.~\ref{sec:Basic} (Appendix~\ref{sec:EigHs}).

\section{\label{sec:Examples}Some illustrative examples}
\subsection{\label{sec:DecoupledSC}Decoupled ``Schr\"odinger cat''}
Here we consider the ``Schr\"odinger cat'' state formed as a coherent superposition of two coherent states:
\begin{equation}
    \left| \psi  \right\rangle  = a_1\left| {{\alpha _1}} \right\rangle  + a_2\left| {{\alpha _2}} \right\rangle
    \label{eqn:sc}
\end{equation}
where each $\left| {{\alpha}} \right\rangle $ is given by Eqn.~\ref{eqn:CS}.

\begin{figure}
    \centering
    \includegraphics{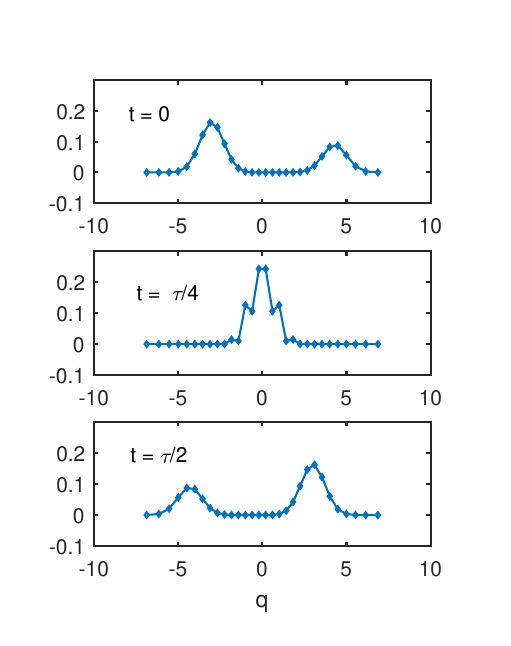}
    \caption{Evolution of a ``Schr\"odinger Cat'' superposition of coherent states (specifics similar to Fig.~\ref{fig:SHOcstestB}). }
    \label{fig:SHOscB}
\end{figure}

\begin{figure}
    \centering
    \includegraphics{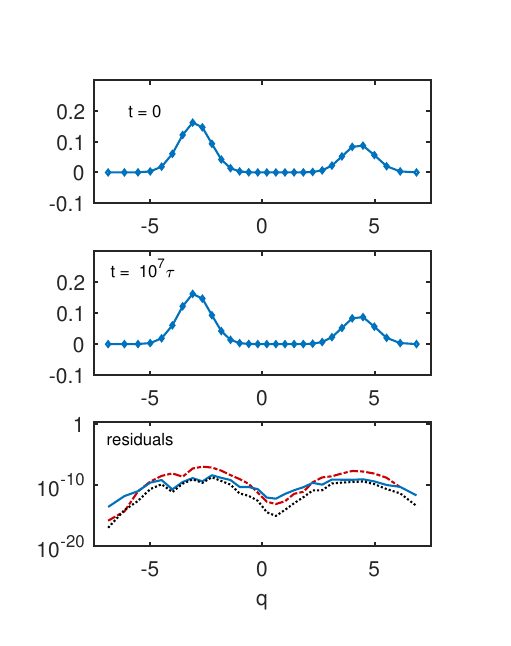}
    \caption{Evolution of a ``Schr\"odinger Cat'' superposition of coherent states (specifics similar to Fig.~\ref{fig:SHOcstest}). }
    \label{fig:SHOsc}
\end{figure}

Figures~\ref{fig:SHOscB} and \ref{fig:SHOsc} are of the same form as figs.~\ref{fig:SHOcstestB} and~\ref{fig:SHOcstest} but showing a state given by Eqn.~\ref{eqn:sc} with $a_1=1/\sqrt{3}$, $\alpha_1=3$, $a_2=\sqrt{2/3}$ and $\alpha_2=-2.1$.  
Again, these are evolved with system-environment interactions turned off. These figures show that the evolution of coherent superpositions is also robust for the ACL model, even though the discrete nature of the truncated SHO shows up in the jagged features of the wavefunction when the two packets collide.
\subsection{\label{sec:Entanglement}Generating entanglement}
\begin{figure}
    \centering
    \includegraphics{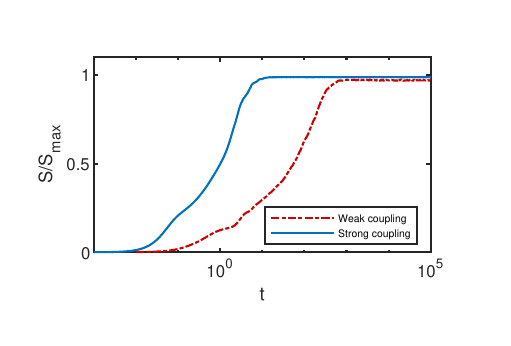}
    \caption{The evolution of the von Neumann entropy for $E_I = 0.03$ (``weak coupling,'' dashed) and $E_I = 0.25$ (``strong coupling,'' solid). Increasing the interaction strength causes the entanglement to increase more rapidly, and also allows the system to come a bit closer to $S_{\max}$. }
    \label{fig:TwoSes}
\end{figure}

Now we consider the case where system-environment interactions are turned on. The interactions will cause an initial product state given by
\begin{equation}
    {\left| \psi  \right\rangle _w} = {\left| \psi  \right\rangle _s}{\left| \psi  \right\rangle _e}
    \label{eqn:ProductState}
\end{equation}
to evolve into an entangled state, where the states of the system and environment are described by the density matrices
\begin{equation}
   {\rho _s} \equiv T{r_e}\left( {{{\left| \psi  \right\rangle }_w}{}_w\left\langle \psi  \right|} \right)
   \label{eqn:RhoSdef}
\end{equation}
and
\begin{equation}
   {\rho _e} \equiv T{r_s}\left( {{{\left| \psi  \right\rangle }_w}{}_w\left\langle \psi  \right|} \right).
   \label{eqn:RhoEdef}
\end{equation}
The Von Neumann entropy,
\begin{equation}
S \equiv tr\left( {{\rho _s}\ln {\rho _s}} \right) = tr\left( {{\rho _e}\ln {\rho _e}} \right),
    \label{eqn:Sdef}
\end{equation}
takes larger values when the degree of entanglement is greater.  The maximum possible value for the entropy is given by 
\begin{equation}
        S_{\max} = \ln \left( {{N_{\min }}} \right)
    \label{eqn:Smaxdef}
\end{equation}
where ${{N_{\min }}}$ is the smaller of $N_s$ and $N_e$.  Figure \ref{fig:TwoSes} shows the evolution of the entropy for two values of $E_I$.  
Throughout this work we use units where $\hbar=\omega_{SHO} = 1$. We also take   $E_e=0.75$ throughout.  For Fig.~\ref{fig:TwoSes} the initial state has the product form (Eqn.~\ref{eqn:ProductState}) with ${\left| \psi  \right\rangle _s}$ given by the Schr\"odinger cat state discussed above and ${\left| \psi  \right\rangle _e}$ given by the 500$^{th}$ eigenstate of $H_e$ (indexed from lowest to highest eigenvalues). The choice of ${\left| \psi  \right\rangle _e}$ will be discussed further  Sect.~\ref{sec:Approach2Eqm}. We consider a ``weak coupling'' ($E_I = 0.03$) case and a ``strong coupling'' ($E_I = 0.25$) case. 

\subsection{\label{sec:Einselection}Einselection}
A generic state for $w$ will be an entangled state with non-trivial density matrices, $\rho$, for system and environment. Thus, it is not surprising that in the interacting case that starts in a product state the entanglement entropy will increase from zero.  This process is generally called decoherence, and it would take place with just about any Hamiltonian for $w$\footnote{
See~\cite{Albrecht:1993hf} for some general reflections quantum coherence and the emergence of entanglement.}. For a randomly chosen $H_w$, one would expect the entanglement entropy to become large and the eigenstates of $\rho_s$ and $\rho_e$ to evolve randomly over time without displaying any regular behavior. 

There is a special case of decoherence called ``einselection'' where the initial state and interactions can be set up to favor a special set of eigenstates for $\rho_s$ called ``pointer states.'' The CL model has been used in many of the pioneering studies of decoherence and einselection. Here we revisit some of these results using the ACL model. 

\begin{figure}
    \centering
    \includegraphics{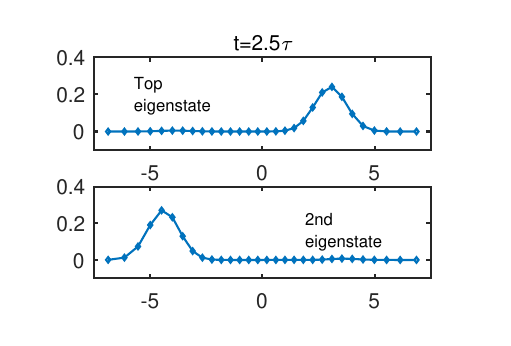}
\caption{The two most probable eigenstates of $\rho_s$ after einselection has completed. The initial states was the Schr\"odinger cat state depicted in Fig.~\ref{fig:SHOcstest}.}
    \label{fig:EinS}
\end{figure}

\begin{figure}
    \centering
    \includegraphics{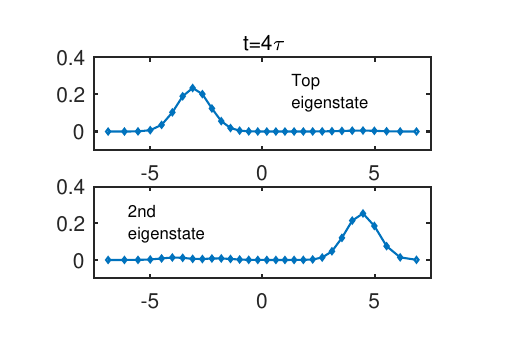}
    \caption{The top post-einselection eigenstates of $\rho_s$ shown in Fig.~\ref{fig:EinS}, but here shown at a different phase in their periodic motion. }
    \label{fig:EinS2}
\end{figure}

The Schr\"odinger cat state depicted in the top panel of Fig.~\ref{fig:SHOsc} is a superposition of two coherent states which can be thought of as ``classical wavepackets.'' Fig.~\ref{fig:EinS} shows what this initial state evolves into by time $t = 2.5\tau$ for the weakly interacting case.
The state of $s$ for $t > 0$ is a density matrix, and Fig.~\ref{fig:EinS} shows the two eigenstates of $\rho_s$ with the largest eigenvalues. One can see that these look like single classical wavepackets.  Figure~\ref{fig:EinS2}
depicts similar information about the state but evolved further in time, to $t = 4\tau$. These eigenstates also look like classical wavepackets, just caught at a different phase of their oscillation.  

There are a variety of technical tools that are useful in studying einselection.  One can anticipate the pointer states and study the decrease in the off diagonal element of $\rho_s$ in that basis (as per~\cite{Zurek1982}). The consistent histories framework can also be useful. The approach we use here, focusing on the eigenstates of $\rho_s$, parallels that developed in~\cite{Albrecht:1992rs} (where a comparison with the consistent histories approach is also presented). We also use the consistent histories method extensively with the ACL model  in~\cite{ACLeqm}.

\begin{figure}
    \centering
    \includegraphics{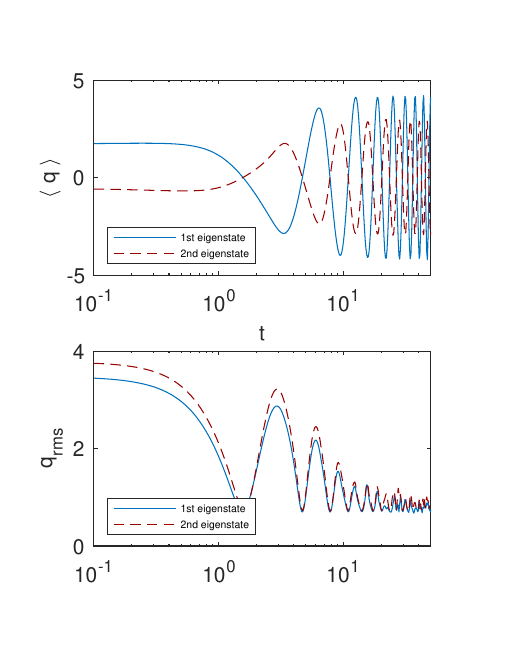}
    \caption{The evolution $\langle q\rangle$ and $q_{rms}$ as a function of time for the top two eigenstates of $\rho_s$ shown in Figs.~\ref{fig:EinS} and~\ref{fig:EinS2}. (The most probable eigenstate is shown with the solid curve, the next most probable is dashed.) One can see these attributes evolve from those of the Schr\"odinger cat initial state (oscillating $q_{rms}$ and small oscillating values of $\langle q\rangle $) to those of individual wavepackets (essentially constant $q_{rms}$ with larger oscillations in $\langle q\rangle$). }
    \label{fig:Moments}
\end{figure}

One can look at this phenomenon a bit more systematically by studying how various moments of the eigenstates evolve over time. Figure~\ref{fig:Moments} shows the time evolution of $\langle q\rangle$ and $q_{rms}$.  One can see how these quantities first exhibit the ``Schr\"odinger cat'' properties, but over time develop the properties of einselected pointer states. 
We conclude that the ACL model nicely reproduces the well-known phenomenon of ``einselection,'' as it should if it is to reflect key properties of the CL model. 

\subsection{\label{sec:Ps}Evolution of the eigenvalues of $\rho$}
\begin{figure}
    \centering
    \includegraphics{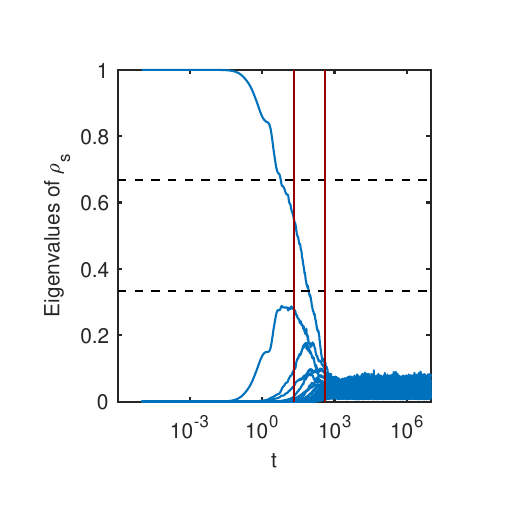}
    \caption{The eigenvalues of $\rho_s$.   The purity of the initial state is reflected in the fact that only one eigenvalue is nonzero initially. The ``einselection time'' (marked by the left vertical line) corresponds to the ``collapse'' of the Schr\"odinger cat pure state into a mixture of wavepackets. The dissipation time (right vertical line) is about $20$ times longer.  The dashed horizontal lines show the probabilities assigned to the two wavepackets in the initial Schr\"odinger cat state.  }
    \label{fig:Ps}
\end{figure}

Figure~\ref{fig:Ps} shows the eigenvalues $p_i$ of $\rho_s$ (for pure states in $w$, the nonzero ones are always identical to the nonzero eigenvalues of $\rho_e$).  The evolution of the $p_i$'s includes the information reflected in the von Neumann entropy (Fig.~\ref{fig:TwoSes}), and clearly shows a transient phase during einselection and a subsequent equilibrium phase where the $p_i$ values are closer together and hold reasonably steady.  One can infer from Fig.~\ref{fig:Moments} that the time to full einselection is $O(20)$.  The dissipation processes that lead to equilibration operate on a time scale roughly $20$ times longer. One can see that by the time einselection is complete there are somewhat more than two nonzero $p_i$'s.  This is related to the relative closeness of the decoherence and dissipation times\footnote{This is in contrast to more macroscopic systems, where the decoherence and dissipation timescales are typically widely separated (see e.g.~\cite{Zurek1986, JoosZeh, schlosshauer2007decoherence}). }, which in turn is connected with the competition between the interaction Hamiltonian (which tries to localize the SHO in space) and the SHO Hamiltonian (which causes localized states to spread).  

\begin{figure}
    \centering
    \includegraphics{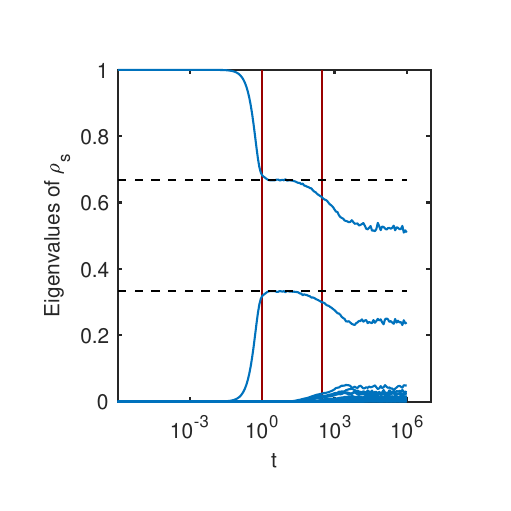}
    \caption{Similar to Fig.~\ref{fig:Ps} but with the initial state and Hamiltonian parameters modified as discussed in the text. This example shows more strongly separated decoherence times and dissipation times. Note in particular that the two top eigenvalues spend an extended period of time at the probability values (dashed lines) assigned to the initial (superposed) wavepackets, indicating that the environment has made a ``good measurement'' of the SHO.}
    \label{fig:PsO}
\end{figure}

Figure~\ref{fig:PsO} shows a case with more widely separated decoherence and dissipation times.  The calculation shown in Fig.~\ref{fig:PsO} uses $\tau_{SHO} = 2\pi\times 10^3$ and the initial state is a superposition of eigenstates of ${\bf q}_{SHO}$ (in the same proportions and locations as the coherent states used in Fig.~\ref{fig:Ps}). These differences mean the interaction term ($\propto {\bf q}_{SHO}$) is not trying to ``chop up'' the initial wavepackets, in contrast to the coherent state initial conditions, which spread across several eigenstates of ${\bf q}_{SHO}$.  

Note that for a while $p_1$ and $p_2$ in Fig.~\ref{fig:PsO} correspond to the probabilities assigned to the wavepackets in the initial superposed state.  
This feature means that the environment can be thought of as ``making a good measuremnt'' of the SHO, in the sense that interactions with the environment have put the SHO in a classical mixture of wavepackets with the right  probabilities.  Later, this good measurement comes unraveled as dissipation sets in. 

\subsection{\label{sec:Copycat}The copycat process}
\begin{figure}
    \centering
    \includegraphics{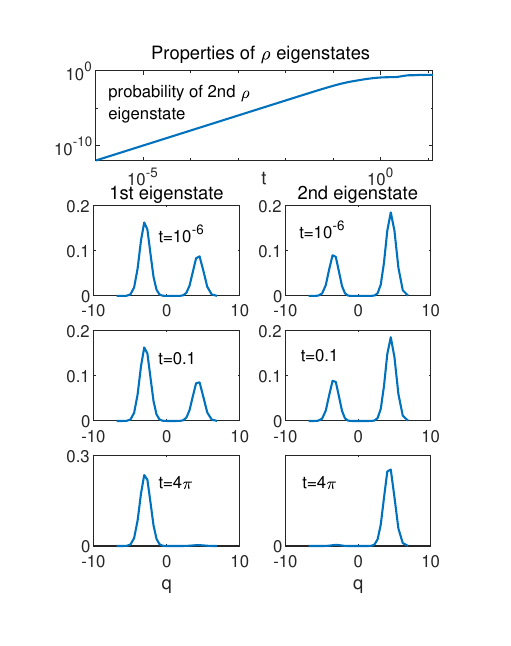}
    \caption{Copycats in the early stages of entanglement: The system is initially taken to be in a Schr\"odinger cat state (2nd row, left panel) which becomes entangled with the environment as it evolves. The 2nd eigenvalue and ${\left| {\psi \left( q \right)} \right|^2}$ for the first two eigenstates of $\rho_s$ are shown from early stages of the evolution.  The 2nd eigenstate generically takes the mirror image ``copycat'' form over several decades of evolution before finally einselecting to a coherent state form. }
    \label{fig:CopyCat}
\end{figure}

Our ACL model allows us to scrutinize the very first steps of the einselection process.  In doing so we've become intrigued by certain aspects of these early stages.  Figure~\ref{fig:CopyCat} shows the early evolution of the 2nd eigenvalue and eigenstate of $\rho_s$, in the case where the system starts in a pure Schr\"odinger cat state which becomes entangled with the environment. The eigenstate takes an intriguing form that appears to be a ``mirror image'' of the initial state, and remains in this form in a transiently stable way over several decades of time evolution (and growth of $p_2$).  We call these mirror image states ``copycat'' states. In~\cite{CopyCat} we systematically investigate this curious behavior and argue that it is quite generic for early time evolution of Schr\"odinger cat states.  We also discuss how this phenomenon generalizes in the case of larger numbers of ``cats.'' 

\section{\label{sec:Approach2Eqm}Approach to Equilibrium}
\begin{figure}
    \centering
    \includegraphics{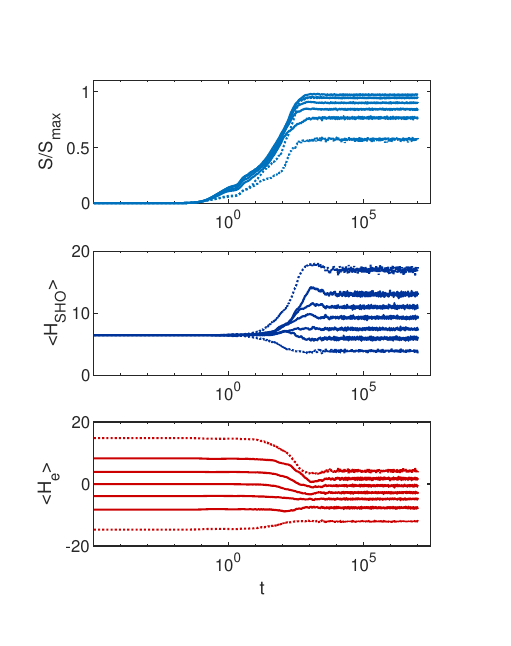}
    \caption{The evolution of entropy and subsystem energies over time, choosing the environment initial state from among the eigenstates of $H_e$. The dotted curves correspond to the very lowest and very highest eigenvalues, and the other curves run from lowest to highest index (from the set given in the text) corresponding to the low or high positions on plots.  Each entropy curve stabilizes over time around its highest value, and the corresponding energy curves stabilize as well (implying no net energy flow after the initial transient).  These are characteristics of equilibration.}
    \label{fig:EQMbE}
\end{figure}

\begin{figure}
    \centering
    \includegraphics{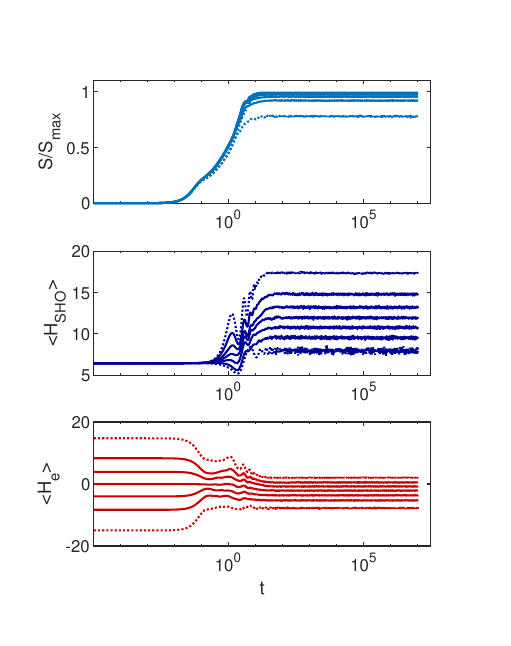}
    \caption{This figure is constructed the same way as Fig.~\ref{fig:EQMbE} except here strong coupling ($E_I=0.25$) is shown. The behavior is broadly similar in terms of equilibration (with the overall entropies tending to be larger, as mentioned with Fig.~\ref{fig:TwoSes}). In the strong coupling case the backreaction tends to significantly impact the effective potential in which the oscillator moves, and can even shift around the location of the minimum. The additional broad oscillations on the approach to equilibrium vs Fig.~\ref{fig:EQMbE} appear to be related to this effect. }
    \label{fig:EQMbES}
\end{figure}

  Figure~\ref{fig:EQMbE} shows the evolution of entropy and energies over time for a variety of initial states of the environment for the weakly coupled case ($E_I = 0.03$).
The strong coupling case is shown in Fig.~\ref{fig:EQMbES}.
 We start the environment in an eigenstate of $H_e$, with values of the index $i_e$ chosen from $\{1,100,200,300,400,500,600\}$ (ordered so the $i_e$ runs from lowest to highest eigenvalues). Each case shows characteristics of equilibration.  Each curve corresponds to a single realization of the random Hamiltonians used in $H_e^I$ and $H_e$.  We have found that the noteworthy features of the curves remain unchanged as different realizations are chosen, except for the cases at the ends of the spectrum where the density of the eigenstates of $H_e$ is low and the noise from the randomness shows up more strongly.  Also note that the timescale for the first significant evolution of the entropy up from zero is similar for all values of $i_e$ except the extremal ones, which rise more slowly.  This also chimes with what one might expect from the low density of states case.  

The finite sizes of the systems makes standard definitions of temperature difficult to utilize.  Still, in~\cite{Albrecht:2022mrx} we have found some generalized notions of equilibration and even thermalization apply, without reference to temperature.  These ideas allow us to understand the behavior of the ACL model as ``equilibration'', as suggested strongly by Figs.~\ref{fig:EQMbE} and~\ref{fig:EQMbES}. 

\section{\label{sec:RCL} The reduced Caldeira-Leggett model}

The ACL can be reduced by replacing the SHO with a single qubit, and turning off the self-Hamiltonians of both the system and the environment.  The resulting ``reduced Caldeira-Legget'' (RCL) model has this Hamiltonian\footnote{The RCL is the same model discussed in~\cite{Albrecht:1992rs} with $H_1^{\uparrow} = - H_1^{\downarrow}$ and $E_1=0$.}:
\begin{equation}
    H_{RCL} = \lambda S_z\otimes H_e^I
    \label{eqn:RCL}
\end{equation}
where ${S_z} \equiv \left|  \uparrow  \right\rangle \left\langle  \uparrow  \right| - \left|  \downarrow  \right\rangle \left\langle  \downarrow  \right|$.
We consider an initial Schr\"odinger cat state of the form
\begin{equation}
   {\left| \psi  \right\rangle _s} = a_1\left|  \uparrow  \right\rangle  + a_2\left|  \downarrow  \right\rangle 
\end{equation}
and present results using $a_1=1/\sqrt{3}$ and $a_2=\sqrt{2/3}$ (as with the SHO Schr\"odinger cat state discussed above). 

\begin{figure}[h]
    \centering
    \includegraphics{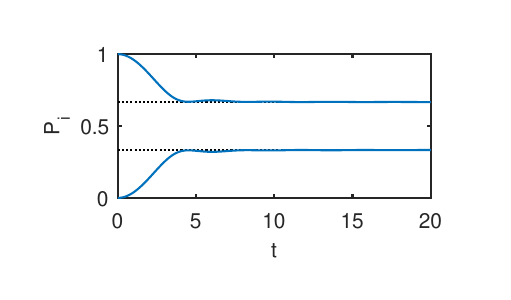}
    \caption{Eigenvalues of $\rho_s$ ($P_i$) as a function of time for the RCL model. The simplified form of the RCL (vs the ACL) model allows the $P_i$'s to settle at the values set by the initial Schr\"odinger cat state, producing a stable ``quantum measurement.'' }
    \label{fig:RCL_Ps}
\end{figure}

Figure~\ref{fig:RCL_Ps} 
shows the evolution of the eigenvalues of $\rho_s$.  The simplified form of the RCL model means there is no self Hamiltonian for the system competing with the interaction term, and the pointer states are simply the spin states $\left\{ {\left|  \uparrow  \right\rangle ,\left|  \downarrow  \right\rangle } \right\}$ determined by the form of the interaction Hamiltonian. Thus the ``good measurement'' behavior (with the $p_i$'s stabilizing at the values ${\left| {{a_1}} \right|^2}$ and ${\left| {{a_2}} \right|^2}$ given by the dotted lines) is realized more robustly than in the case depicted in Fig.~\ref{fig:PsO}.  

\begin{figure}[h]
    \centering
    \includegraphics{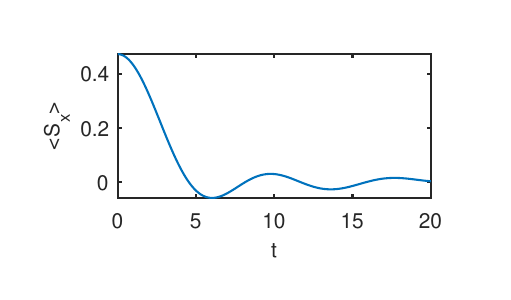}
    \caption{ The quantity $\langle S_x \rangle$, giving the off diagonal elements of $\rho_s$ in the spin basis for the RCL model.  While the the spin basis is nominally the pointer basis, the inefficiencies of einselection in the RCL model allow significant deviations from zero at late times. }
    \label{fig:RCL_Sx}
\end{figure}

\begin{figure}
    \centering
    \includegraphics{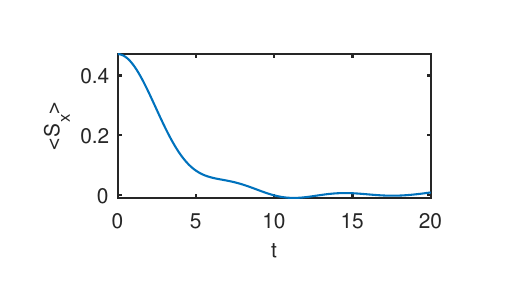}
    \caption{\label{fig:RCLx_Sx}Off diagonal elements of $\rho_s$ in the spin basis for the RCL model amended to include a self-Hamiltonian for the environment.  As discussed in the text, this modification suppresses the late time oscillations observed in Fig.~\ref{fig:RCL_Sx}. (The added term in $H_w$ has the form of the last term in Eqn.~\ref{eqn:Hform}, with $H_e$ defined by Eqn~\ref{eqn:Hedef} with $E_e=0.025$ and $E_e^0=0$.)  }
\end{figure}

Figure~\ref{fig:RCL_Sx} 
shows the (real parts of the) off-diagonal elements of $\rho_s$ in the pointer state basis (a.k.a. $\left\langle {{S_x}} \right\rangle $). From this perspective, the approach of $\left\langle {{S_x}} \right\rangle $ toward zero reflects the process of einselection.  The uneven fluctuations in the approach toward zero reflect inefficiency in the decoherence process.  The RCL model has no self-Hamiltonian for $e$ and thus the decoherence boosting effects depicted in Fig.~\ref{fig:HeHI} are not available (Figure~\ref{fig:RCLx_Sx} shows results comparable to Fig.~\ref{fig:RCL_Sx} but with a self-Hamiltonian added, and one can see that the oscillations have essentially disappeared).  In Sec.~\ref{sec:NMR_Loschmidt} we discuss how such curves relate to phenomena seen in NMR experiments, and connect these features with a phenomenon known as ``Loschmidt echos.''  And in~\cite{CopyCat} we explore more systematically the variety of behaviors possible for the full complex values of the off-diagonal elements of $\rho_s$.

\section{\label{sec:Comparison}Comparison with other work}
\subsection{Limits of einselection} As reviewed in~\cite{schlosshauer2007decoherence}, Zurek and collaborators have (in the context of CL models) considered various interesting limits which cause different pointer states to be selected by the decoherence processes. We have reproduced each of these limits in this paper. 

The ``quantum measurement limit'' occurs when the interaction term dominates.  In that limit the pointer states are eigenstates of the interaction Hamiltonian.  The case we illustrate in Fig.~\ref{fig:PsO} is approaching the quantum measurement limit. Another extreme is the ``quantum limit,'' where the self-Hamiltonian of the system dominates.  The pointer states in this case are the energy eigenstates of the system. We explore this limit for the ACL model in Appendix~\ref{sec:QuntLim}.

When the effects of the interaction term and self-Hamiltonian are similar (the ``intermediary regime''), the pointer states tend to be the coherent states.  Much of our discussion in Sect.~\ref{sec:Examples} covers this regime\footnote{For the way we have parameterized $H_e^I$, the environment size $N_e$ impacts the strength of the interaction term.  When that effect is taken into account, the effective strengths of $H_e^I$ and $H_s$ are similar for the ``weakly interacting'' parameters chosen in the first parts of Sect.~\ref{sec:Examples}.}.

\subsection{\label{sec:deviations}Other treatments of the CL Model and the  Markovian limit}

Physicists studying decoherence and einselection often encounter the CL model in the context of master equations.  These master equations describe the evolution of the system density matrix, without the need to specify the full dynamics of the surrounding environment~\cite{schlosshauer2007decoherence}. However, to derive such master equations, approximations such as the ``Born approximation'' and ``Markov approximation" are typically made.  Both are reviewed in~\cite{schlosshauer2007decoherence, BreuerPetruccione}, but we draw attention to the Markov approximation here. 

In the Markov approximation, the environment is assumed to be `memoryless.' This assumes any quantum correlations between parts of the environment that were created due to system-environment interactions are quickly `forgotten.' `Quickly forgotten' is often quantified by the statement $\tau_{corr} \ll \tau_{s}$---where $\tau_{corr}$ is the timescale for destroying such dynamically generated environmental self-correlations and $\tau_{s}$ is the timescale over which the system density matrix changes noticeably~\cite{schlosshauer2007decoherence}. The Markov approximation is often appropriate for cases where the system and environment are weakly coupled, for example. However, there are many situations of physical interest where this inequality does not hold and the influence of environmental correlations on the system cannot be neglected~\cite{schlosshauer2007decoherence,BreuerPetruccione}.  Nevertheless, the Markov approximation is standard in many master equation approaches to studying decoherence, including the CL master equation---though there are exceptions, e.g.~\cite{PhysRevD.45.2843}.  

Other assumptions that typically enter into deriving the CL master equation are a high temperature environment---such that the thermal energy of the environment is much larger than the energy scale set by the system’s natural frequency---and an environment which is described by an `ohmic’ spectral density with a suitable UV cutoff scale.

One consequence of these assumptions, along with the Markov approximation, is that the CL master equation typically predicts exponential decay for the off-diagonal elements of the system density matrix---an exponential rate of decoherence.  This exponential result is also found in other parts of the literature on decoherence, such as scattering induced decoherence~\cite{JoosZeh,schlosshauer2007decoherence} and particular limits of spin-boson models~\cite{Unruh:1994az,schlosshauer2007decoherence}.  While even within master equation approaches it is known that exponential decay is not always valid~\cite{Anglin1997, BreuerPetruccione,schlosshauer2007decoherence}, there remains strong focus in the literature on exponential decay.

In contrast to the CL master equation approach, our results from the ACL and RCL models show a more varied range of time dependence in the decay of off-diagonal system density matrix elements. Examining  Figs.~\ref{fig:RCL_Sx} and~\ref{fig:RCLx_Sx}, for example, the decay is not exponential at all (except perhaps in a narrow time range). In our work we have not made any assumptions of Markovian evolution, we have simply solved the Schr\"odinger equation directly for system and environment in its fully unitary form as discussed in Sec.~\ref{sec:Basic}. Therefore, deviations from Markovian behavior and exponential decay should be unsurprising. As Zurek and collaborators~\cite{Zurek1982,CPZ1,CPZ2} explicitly note in the context of the formalisms they develop---which have some parallels to our work---exponential behavior is a very special case.

Furthermore, our main motivation for developing the ACL model is to study equilibrium systems. The detailed balance exhibited by such systems would imply that the ``forgetting'' of correlations and ``(re-)emergence'' of correlations should contribute equally to the physics. Markovian treatments are by construction unable to include such features.

\subsection{\label{sec:NMR_Loschmidt}Loschmidt echos and NMR}
In~\cite{CPZ1, CPZ2}  Cucchietti, Paz, and Zurek (CPZ) consider a model very similar to our RCL model. They observe oscillations similar to those that appear at later times in our Figs.~\ref{fig:RCL_Sx} and~\ref{fig:RCLx_Sx}.  CPZ point out that these oscillations can be thought of as ``Loschmidt echos,'' and also notes that such features appear in NMR experiments (e.g.~\cite{doi:10.1063/1.475664}).  The notion of Loschmidt echo originates in discussions of fluctuations in the arrow of time (the direction of entropy increase) in equilibrium systems~\cite{Goussev:2012}.  The Loschmidt echo refers to the possibility of \emph{partial} time reversal occurring. CPZ note that in expressions like our Eqn.~\ref{eqn:RCL}, $H_e^I$ multiplies  $\left|  \uparrow  \right\rangle \left\langle  \uparrow  \right| $ and $\left|  \downarrow  \right\rangle \left\langle  \downarrow  \right|$ with an opposite sign, something that can be thought of as effectively generating two evolutions in the $e$ subspace, each the time reverse of the other. In this way they make the connection with Loschmidt echos. 

In this paper we have interpreted the oscillations as inefficiencies (or more specifically, non-monotonicity) in the establishment of entanglement between system and environment.  These inefficiencies reflect the finite environment size and various properties of $H_w$, as discussed in Sect.~\ref{sec:RCL}.  This narrative also seems to work well for the NMR results, where it appears that in the cases where the oscillations occur the environment is effectively finite (comprised predominantly of nearby spins).  While the different narratives (``inefficient decoherence'' and ``partial time reversal'') may superficially sound quite different, in this case they are describing the same phenomenon. 

\section{\label{sec:Conclusions}Conclusions}
We have presented a modified version of the classic Caldeira-Leggett (CL) model which can be studied using full unitary evolution in the combined system-environment space. This adapted Caldeira-Leggett (ACL) model enables explorations beyond the various approximation schemes which are usually used with the CL model.   Examples of such new explorations are presented in companion papers devoted to studying whether the notion of einselection makes sense under conditions which do not exhibit an arrow of time~\cite{ACLeqm}, and examining the very earliest stages of the einselection process~\cite{CopyCat}. This paper provides background information, including details of how the ACL model is constructed and of our highly accurate numerical techniques.

We have also reproduced a number of well-known results from the literature on decoherence and einselection.  These build our confidence that the ACL model is well suited for our intended studies, and also help us know its limitations.  Our full numerical treatment enables detailed scrutiny of all aspects of the process of einselection, and our extensive graphical representations of that phenomenon may provide a useful resource for those wishing to learn more about einselection. 

In addition, Sec.~\ref{sec:Copycat} briefly introduces new results which anticipate the work presented in~\cite{CopyCat}.  Also, experts versed in the notion of the ``quantum limit'' of the einselection process might enjoy our exploration of that limit in Appendix~\ref{sec:QuntLim}.  While such experts would not find those results altogether surprising, we appreciate the way the ACL model allows us to explore interesting intermediate behaviors on the way to the full quantum limit. 

We conclude that the ACL model provides a reliable tool with which to explore decoherence and einselection under conditions which cannot be treated using the standard approximation schemes. 

\section{\label{sec:Acknowlegments}Acknowledgments}
We thank Wojciech Zurek for numerous inspiring conversations (over many years in the case of one of us, AA) which created the foundation for this work. We also thank Fabio Anza, Nick Curro and Zhipang Wang for discussions of NMR phenomena.   This work was supported in part by the U.S. Department of Energy, Office of Science, Office of High Energy Physics QuantISED program under Contract No. KA2401032. 

\vspace{1cm}

\appendix
\section{\label{sec:QuntLim}The Quantum Limit}
In~\cite{Paz:1998ib} Paz and Zurek consider the case where $H_s$ dominates over the other terms in $H_w$. They call this case the ``quantum limit.''  We consider the quantum limit in the context of the ACL model here. 
While our results are broadly consistent with the existing literature, we also noticed several interesting behaviors which have so far not been reported. 

The pointer states in the quantum limit have been shown to be the eigenstates of $H_s$~\cite{Paz:1998ib}.  To explore this limit with the ACL model we use the ``predictability sieve'' ideas~\cite{Zurek1993predictsieve,ZurekTimeAsymm,Zurek:1992mv, PhysRevA.53.655, PhysRevE.50.2538, Dalvit_2005,RevModPhys.75.715, schlosshauer2007decoherence}, which are grounded in the notion that the pointer states should be the states which are most stable against entanglement with the environment\footnote{While we're not doing a thorough sifting of the entire Hilbert space in our analysis here, we find utilizing ``predictability sieve'' arguments to make comparisons between specific states sufficient for our purposes.}.  Here we consider the case where $E_I=3\times 10^{-3}$ and $E_e= 0.015$, well below the values considered elsewhere in this paper, while keeping $H_s$ the same. We considered initial states of product form (Eqn.~\ref{eqn:ProductState}) where $\left| \psi  \right\rangle _s$ is either an eigenstate of $H_s$, the Schr\"odinger cat (SC) state shown in Fig.~\ref{fig:SHOsc}, or a single ($\alpha = 3$) coherent state (CS), and compare the evolution in these cases. 

\begin{figure}
    \centering
    \includegraphics{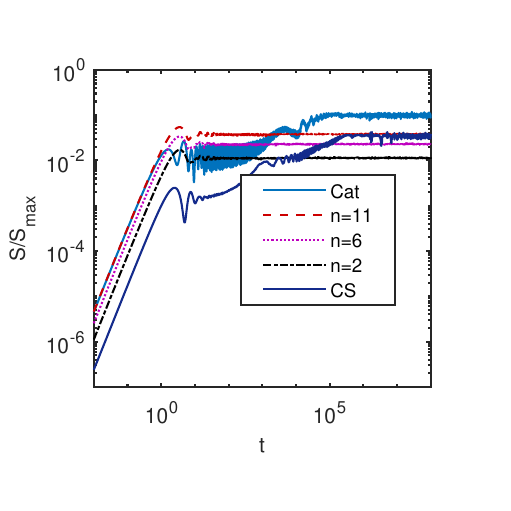}
    \caption{von Neumann Entropy evolution in a case where $H_s$ dominates.  The initial state is a product state with $\left| \psi  \right\rangle _s$ given by a cat state (solid, upper), energy eigenstates with index 11 (dashed), 6 (dotted), 2 (dot dashed) or a single coherent state (CS, solid, lower). In the idealized ``quantum limit'' where $H_s$ fully dominates, the energy eigenstates are the pointer states which are expected to be the ``most robust'' against the onset of entanglement. In this example we see that there is an early and intermediate period where the coherent state is favored, and it is only later that the full einselection of the energy eigenstates sets in.}
    \label{fig:QL_Ss}
\end{figure}

\begin{figure}
    \centering
    \includegraphics{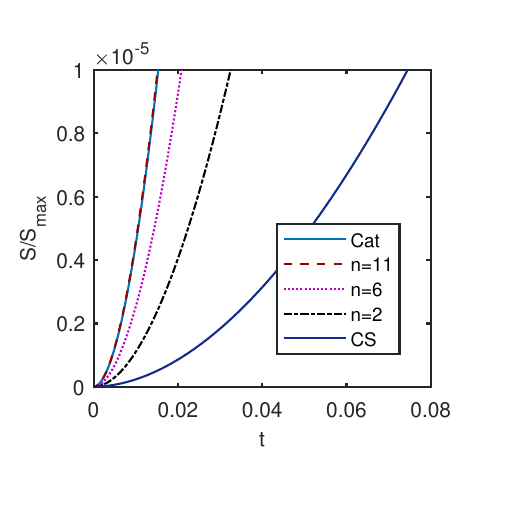}
    \caption{Zooming in on Fig.~\ref{fig:QL_Ss} and showing linear axes.  The different initial rates of the onset of entanglement are clearly exhibited here.  In this initial period the coherent state  (CS) is the most robust against entanglement.}
    \label{fig:QL_SsZoomLin}
\end{figure}

Figures~\ref{fig:QL_Ss} and~\ref{fig:QL_SsZoomLin} show the evolution of the von Neumann entropy for these choices of initial state.
Identifying robustness against entanglement with small values of the entropy at late times, once can conclude that the cat state is least robust, the lower $n$ energy eigenstates are most robust, and the coherent state comes in about the same as $n=11$. (We found the larger $n$ values reach larger late-time entropies but, as discussed in Appendix~\ref{sec:EigHs}, we also expect significant finite size effects to come in for the higher eigenstates of $H_s$.) Interestingly, the cat and the CS states exhibit much lower entropies for several decades of earlier time evolution which suggests a different (and transient) hierarchy of robustness.  Furthermore, if one uses the timescale for the early time onset of entanglement as the measure of robustness, the coherent state is significantly more robust than the other cases considered.  The original work on the quantum limit~\cite{Paz:1998ib} only showed the stability of eigenstates of $H_s$ at late times, and did not actually compare the rate of onset of entanglement.  It appears that during the early and intermediate periods the coherent states exhibit the strongest resistance to entanglement (reflecting the sort of behavior demonstrated in Sect.~\ref{sec:Copycat}), and only later does the long time behavior set in ultimately favoring the energy eigenstates. 

\begin{figure}
    \centering
    \includegraphics{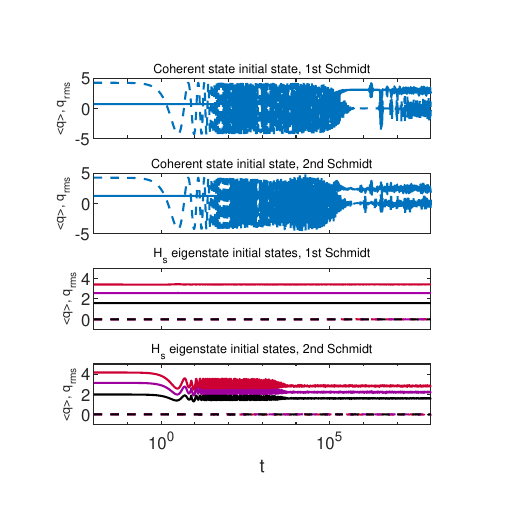}
    \caption{Evolution of $\langle q \rangle$ (dashed) and $q_{rms}$ (solid) for the first and 2nd most probable eigenstates of $\rho_s$ starting with different initial states. The coherent state initial state in the top panel initially exhibits the usual oscillatory behavior, but then degrades into noise. The energy eigenstate initial state in the third panel is highly stable as expected in the quantum limit. The 2nd Schmidt states (second and fourth panels) are ill defined at $t=0$, but they emerge due to the interactions with the environment.  Each roughly reflects the behaviors of their corresponding 1st Schmidt, although the energy eigenstate initial state case takes a while to get there. The energy eigenstate initial states, in order descending from the top curve are $n=11$, $n=6$, and $n=2$.  }
    \label{fig:QL_qInfo}
\end{figure}

Figure~\ref{fig:QL_qInfo} shows the evolving properties of the top two eigenstates of $\rho_s$ (aka ``Schmidt states'').  
For the coherent initial state (CS), these Schmidt states exhibit the properties of coherent states (steadily oscillating  $\langle q \rangle$ and constant $q_{rms}$) for an extended period before degrading into more noisy, unstable behavior.  This fits with the narrative we surmised from the entropy curves. For the energy eigenstate initial state the top Schmidt is perfectly stable, maintaining the energy eigenstate features, as expected for a pointer state. The second Schmidts (panels 2 and 4) emerge due to the process of decoherence (they are ill defined at $t=0$, where $\rho_s$ has only one nonzero eigenvalue) and reflect interesting properties of the decoherence process (also discussed in Sect.~\ref{sec:Copycat}).  For the energy eigenstate initial state, the 2nd Schmidt (4th panel) first reflects some oscillating behavior before becoming highly stable as well. The curves for CS initial state case exhibit a transient period of stable behavior around $t=10^6$ but the stability does not extend to other moments of the Schmidts.  Those Schmidts are not actual eigenstates of $H_s$.

\begin{figure}
    \centering
    \includegraphics{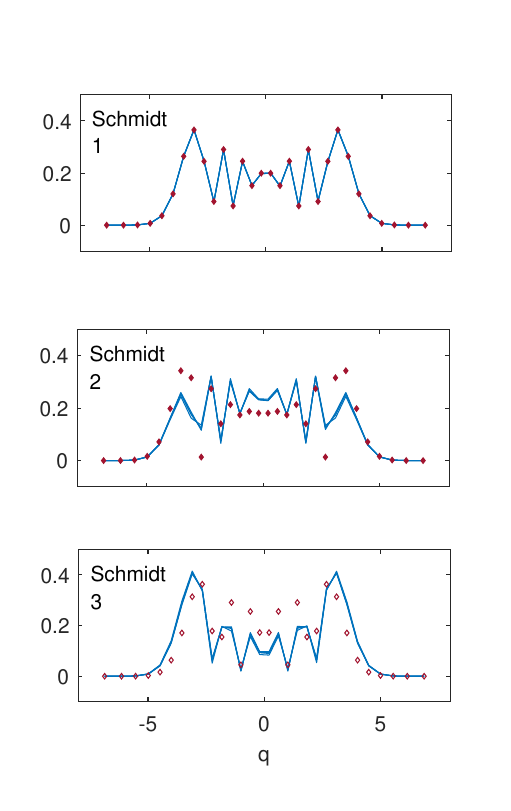}
    \caption{Energy eigenstates as pointer states: Snapshots of the $|\psi(q)|$ for the three most probable eigenstates of $\rho_s$ (solid curves). Each panel shows the state at $t=10^4$, $t=10^5$, $t=10^6$ and $t=10^7$.  These correspond to the period of time where all the curves in the 4th panel of Fig.~\ref{fig:QL_qInfo} are very stable. The wavefunctions at these different times are mostly indistinguishable to the eye, indicating that the stability goes well beyond the two moments plotted in Fig.~\ref{fig:QL_qInfo}.  Also plotted on each panel are (top to bottom) the $n=6$, $n=7$ and $n=5$ eigenstates of $H_s$ (markers).  As discussed in the text, the behaviors depicted here strongly reflect the fact, developed in earlier literature, that the energy eigenstates of $H_s$ are the pointer states in the quantum limit. We are especially intrigued by the 2nd and 3rd panels which illustrate that Schmidt states similar to these pointer states are distilled out of the messy physics of decoherence by the einselection process. (The eigenvalues are $0.98$, $0.015$ and $0.004$.)}
    \label{fig:QL_EEsnaps}
\end{figure}

Figure~\ref{fig:QL_EEsnaps} shows the full wavefunctions of the Schmidt states for the case where the initial state is an energy eigenstate (the top two of these have moments shown in Fig.~\ref{fig:QL_qInfo}). These ``snapshots'' are taken for $t>=10^4$, where the corresponding curves in in Fig.~\ref{fig:QL_qInfo} are very stable. 
One can see that these Schmidts are highly stable in this time period and are very close to true eigenstates of $H_s$\footnote{It is interesting that despite their high degree of stability, the 2nd and 3rd Schmidts do not match perfectly to eigenstates of $H_s$. We conjecture that this is due to a small ``effective potential'' for the SHO due to the interactions with the environment.}.

\begin{figure}
    \centering
    \includegraphics{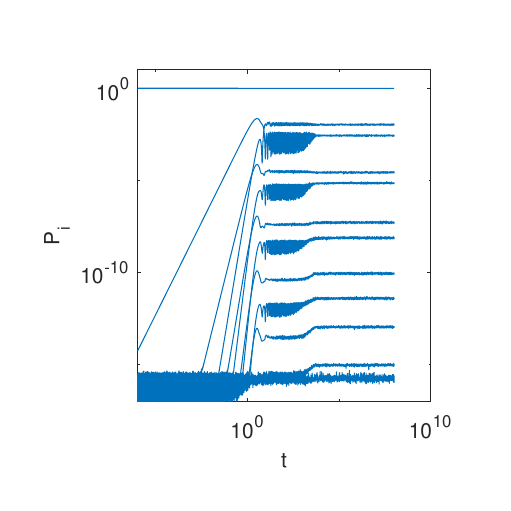}
    \caption{The evolution of the top 12 eigenvalues of $\rho_2$ for the case where the system starts in its $n=6$ energy eigenstate. The interesting crossing behavior and alternating ``noise buldges'' are discussed in the text.}
    \label{fig:QL_Ps}
\end{figure}

Finally, in Fig~\ref{fig:QL_Ps} we show the evolution several of the top eigenvalues of $\rho_s$.  
Not surprising for a case with very weak interactions, the top eigenvalue does not deviate too far from unity. We also note the interesting ``crossover'' behavior, where alternate eigenvalues rise faster and experience an initial noisy period in equilibrium before settling down.  We speculate that this behavior is related to the eventual emergence of the other eigenstates of $H_s$ as eigenstates of $\rho_s$ and suspect that the two types of behavior are related to the parity of the energy eigenstates that emerge. 

All the results reported in this Appendix appear to be consistent with statements in the literature about the quantum limit case, although we've not done a sufficiently thorough investigation to explicitly demonstrate that eigenstates of $H_s$ are the {\em most robust} against interactions with the environment out of all possible choices.  The behavior of the other eigenstates of $\rho_s$ noted here is intriguing.  While it appears broadly consistent with established ideas about the quantum limit, we've not found any report of these particular effects in the literature. 

\section{\label{sec:EigHs} Eigenstates of $H_s$}
Our form of $H_s$ does a nice job of describing the evolution we associate with the continuum SHO using a finite Hilbert space, as discussed in the body of this paper.  Here we provide some further information, focusing especially on the eigenstates of $H_s$.

\begin{figure}
    \centering
    \includegraphics{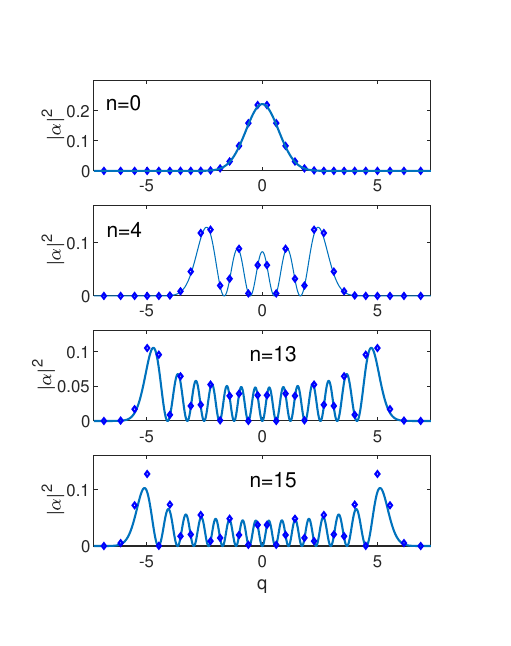}
    \caption{Energy eigenstates of the truncated SHO (markers) along with the corresponding continuum SHO eigenstates (curves). The two track one another nicely, although the tracking comes under a bit of strain for the $n=15$ state where the continuum state starts pressing up against the finite bounds on $q$ which exist in the truncated case.}
    \label{fig:EigsHsSet1}
\end{figure}

\begin{figure}
    \centering
    \includegraphics{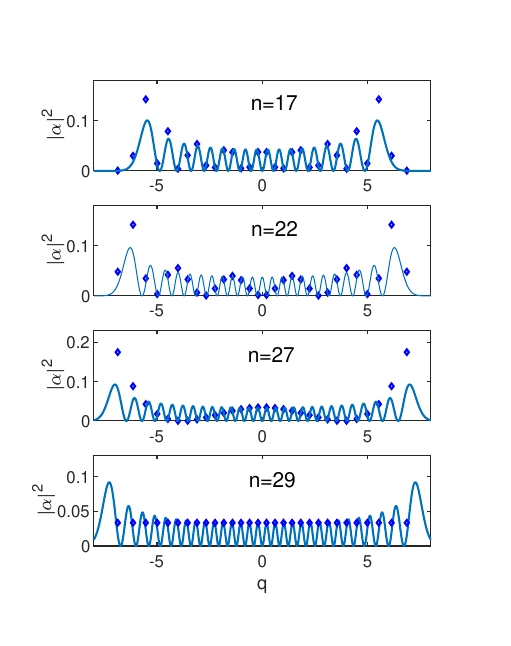}
    \caption{Energy eigenstates of the truncated SHO (markers) along with the corresponding continuum SHO eigenstates (curves) shown for larger $n$ values. The tracking behavior noted in Fig.~\ref{fig:EigsHsSet1} is present here as well, although the edge effects are more pronounced. For these $n$ values, taken alone the markers appear to trace very different curves, but this is only because the discrete grid on which they lie beats in an interesting way off of the frequencies exhibited by the continuum states. }
    \label{fig:EigsHsSet2}
\end{figure}

Figure~\ref{fig:EigsHsSet1} and~\ref{fig:EigsHsSet2} depict selected eigenstates of $H_s$ shown along with their continuum counterparts, given in the $q$ basis. 
In these figures the states of the truncated SHO are shown only as markers (with no connecting lines) to emphasize the fact that these states exists in a finite space. (In these figures the normalization is adjusted for easy cross-comparison.) One can see that the lower energy eigenstates (Fig.~\ref{fig:EigsHsSet1}) follow the behavior of the continuum states quite nicely. As one approaches higher energies (Fig.~\ref{fig:EigsHsSet2}) the eigenstates reach the edge of the finite $q$ range and start showing nonzero values at the $q$ edges.  This leads to behaviors at high energies that deviate significantly from the details of the continuum case, although some broad features remain.  Because of this behavior, we have avoided studying cases that put the SHO in higher energy excitations in this paper as well as in other work using the ACL model, since our intention is to represent a realistic SHO as well as possible.  We found for example that coherent states with considerably higher amplitudes than those shown here executed interesting combinations of reflection and periodic transmission at the $q$ boundaries, hardly surprising given the forms of the higher energy eigenstates.

\begin{figure}
    \centering
    \includegraphics{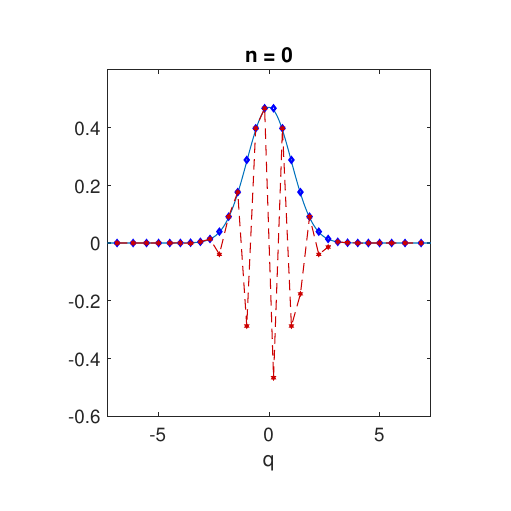}
    \caption{ The ground state of $H_s$. Blue: ${\left| {\psi \left( q \right)} \right|}$, Red: $Re\left( {\psi \left( q \right)} \right)$. The state is defined in an $N_s$ dimensional Hilbert space, and the discrete nature of that space is expressed by the markers on the plot. The markers are connected by lines in order to reference the continuum SHO case.  In the case of  ${\left| {\psi \left( q \right)} \right|}$ this correspondence appears to be simple, but $Re\left( {\psi \left( q \right)} \right)$ has jagged features not found in the continuum SHO ground state.  We discuss the nature of these features in the text and note that while appearing to be exotic, they do not interfere with an intuitive understanding of our truncated SHO, which overall exhibits behaviors very similar to the continuum case.   }
    \label{fig:GSwig}
\end{figure}

We also note an exotic feature that appears as an artifact of our finite construction. Figure~\ref{fig:GSwig} shows the same ground state wavefunction shown in the top panel of Fig.~\ref{fig:EigsHsSet1}, but here we show $\psi(q)$ both with and without the norm.
The un-normed values show a jaggedly varying sign.  In continuum terms such jaggedness would result in an energy much higher than the ground state energy, but our $H_s$ has correspondingly complicated off diagonal elements coupling certain neighboring points which make the $\psi(q)$ shown truly the lowest energy state.  We've also checked that these considerations do not disrupt our use of continuum intuition with other eigenstates of $H_s$, at least for $n\lesssim N_s/2$.  The robust behavior of the isolated oscillator reported in Figs.~\ref{fig:SHOcstest} and~\ref{fig:SHOsc} also supports our confidence that our truncated SHO is overall a good approximation to the continuum case.

\section{\label{sec:spectra}Energy spectra}
\begin{figure}
    \centering
    \includegraphics{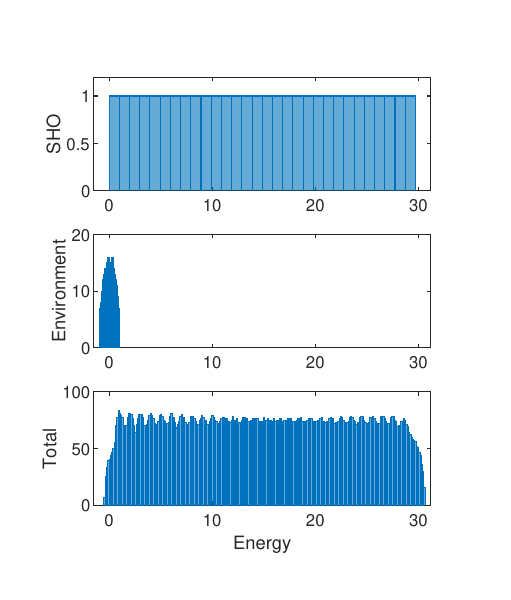}
    \vspace{-0.8cm}
    \caption{The eigenvalues spectra of $H_w$ (lower panel) and its two main components, the SHO (upper) and $H_e$ (middle). We discuss in the text how these spectra relate to one another and reflect the way the different $H$'s are defined.}
    \label{fig:Ehist}
\end{figure}

Here we take a look at the eigenvalue spectrum of $H_w$, and see how it relates to the spectra of $H_s$ and $H_e$.  Figure~\ref{fig:Ehist} shows histograms of the eigenvalues of each of these $H$'s using $E^e_I = 0.01$, $E_e =  0.05$, $E^0_I = E^e_I$ and $E^0_e=E^e$.  (These are different from the values used in this paper but match those used in~\cite{ACLeqm}, where the spectrum of $H_w$ will be relevant for a discussion of our ``eigenstate einselection hypothesis.'')
The spectrum of a true SHO is flat, and so is the spectrum for our SHO shown in the top panel of Fig.~\ref{fig:Ehist}, although this spectrum is truncated at $E=29$ reflecting the finite Hilbert space inhabited by our truncated SHO. The spectrum of $H_e$ (middle panel) reflects the well-known ``Wigner semicircle'' property of random matrices.  The eigenvalues of $H_w$ are essentially sums of eigenvalues of $H_s$ and $H_e$ (with a small additional contribution from the interaction term). So it is not surprising that the full spectrum of $H_w$ (lower panel) appears to be a combination of the spectra shown in the upper and middle panels. For these parameters the energy of the SHO dominates, and the spectrum of $H_w$ roughly takes the form of the SHO spectrum (modulated by little semicircles).  For cases where $H_e$ dominates the spectrum of $H_w$ looks more like a single semicircle, with ``wings'' giving a broadening induced by the SHO spectrum. 

\section{\label{sec:Numerical}Numerical techniques and tolerances}
The total Hamiltonian ($H_w$) was constructed as described in the text and then diagonalized numerically.  The initial states were constructed in the appropriate subsystem bases and then expanded in the basis of eigenstates of $H_w$.  Time evolution was performed by rotating the phases of the coefficients of the eigenstates of $H_w$ according to the Schr\"odinger equation. Density matrices for subsystems $s$ and $e$ at a given time were generated by rotating into an $s\times e$ product basis and tracing over $e$ and $s$ respectively. These density matrices were then used to extract information about the two subsystems. (Note, the state of $w$ expressed in the eigenbasis of $H_w$ was always saved so there was never a need to ``rotate back'' and thus no associated noise introduced in the evolution.) Algorithm~\ref{alg:ACL} shows a schematic of the procedure to generate $\rho_s(t)$ and $\rho_e(t)$ in the ACL model. 

\begin{figure}
\begin{algorithm}[H]
\caption{Steps to generate $\rho_{s}(t)$ and $\rho_{e}(t)$ \\ in the ACL model}
\label{alg:ACL}
\begin{algorithmic}
\State \textbf{Inputs:} Choices for $E_{I}$ and $E_{e}$ (overall energy scales of the interaction and environmental Hamiltonian), ${\left| \psi(t=0) \right\rangle _s}$, ${\left| \psi(t=0)  \right\rangle _e}$, $N_{s}$ and $N_{e}$ (system and environment dimensions, set to $N_{s} = 30$ and $N_e = 600$ in the text).
\State {}
\State \textbf{Outputs:} $\rho_{s}(t_{f})$, $\rho_{e}(t_{f})$.
\State {}
\State \textbf{Runtime:} $O(2)$ hrs for all steps, given $N_{s} = 30$, $N_e = 600$, and the computing setup discussed in this appendix.
\State {}
\State \textbf{Procedure:}
\begin{enumerate}
    \item Construct $H_{w}$
        \item[] \vspace{-0.4cm}
        \begin{align} & H_{w} =  H_s \otimes {\bf{1}}_{}^e + {q_s}\otimes H_e^I + {{\bf{1}}^s} \otimes {H_e} \notag \\
        & H_s = {\bf{\hat a}^\dagger}{\bf{\hat a}} + \frac{1}{2} \notag \\
        &  \left\langle n-1 \right|{\bf{\hat a}}\left| n \right\rangle = \sqrt{n}, \quad n \in \{1, N_{s} \}, \quad  {\bf{\hat a}^\dagger} = ({\bf{\hat a}})^\dagger \notag \\
        & {q_s} = {\bf{\hat q}} = \frac{1}{\sqrt{2}} \Big( {\bf{\hat a}} + {\bf{\hat a}^\dagger} \Big) \notag \\
        & H_{e}^{I} = E_{I}R_{I}^{e}, \quad H_{e} = E_{e}R^{e} \notag
        \end{align}
        where $R_{I}^{e}$ and $R^{e}$ are separately constructed random $N_e \times N_e$ Hermitian matrices (see Sect.~\ref{sec:Hes}) 
    \item Diagonalize $H_{w}$
    \item [] Find eigenstates and eigenvalues of $H_{w}$
    \item Construct ${\left| \psi (t=0)  \right\rangle _w} = {\left| \psi(t=0) \right\rangle _s} \otimes {\left| \psi(t=0) \right\rangle _e}$
    \item Expand ${\left| \psi (t=0)  \right\rangle _w}$ in eigenstates of $H_{w}$
    \item[] \vspace{-0.4cm}
    \begin{align}
    {\left| \psi (t=0)  \right\rangle _w} = \beta_{i}{\left| E_{i}  \right\rangle _w} \notag
    \end{align}
    \item Evolve ${\left| \psi \right\rangle _w}$ to desired $t_{f}$
     \item[] \vspace{-0.4cm}
    \begin{align}
    \beta_{i}(t_f) = e^{-i E_{i} t_f}\beta_{i}(t=0) \notag
    \end{align}
    \item Calculate $\rho_{s}(t_{f})$, $\rho_{e}(t_{f})$
     \item[] \vspace{-0.4cm}
    \begin{align}
     & {\rho _s}(t_f) \equiv T{r_e}\left( {{{\left| \psi(t_f) \right\rangle }_w}{}_w\left\langle \psi(t_f)  \right|} \right) \notag \\
     & {\rho _e}(t_f) \equiv T{r_s}\left( {{{\left| \psi(t_f)  \right\rangle }_w}{}_w\left\langle \psi(t_f)  \right|} \right) \notag
    \end{align}
\end{enumerate}
\end{algorithmic}
\end{algorithm}
\end{figure}

\begin{figure}
    \centering
    \includegraphics{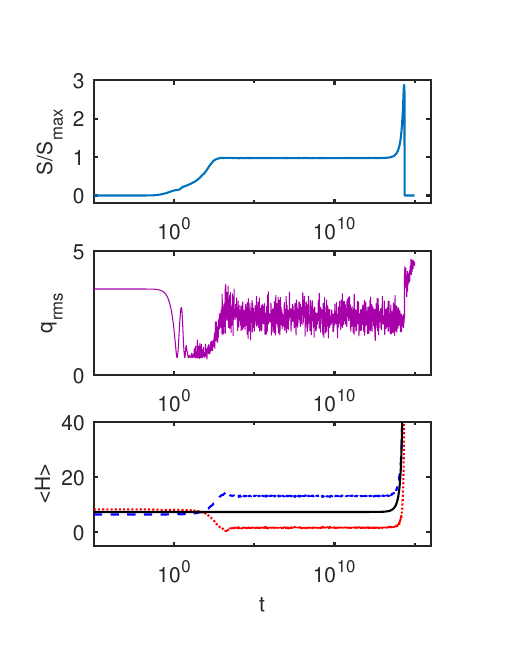}
    \caption{Various quantities are shown evolved over a huge time range to illustrate the point where our numerical computations fail. Top panel: Entropy. Middle panel: $q_{rms}$ of the most probable eigenstate of $\rho_s$ (discussed in Fig.~\ref{fig:Moments}). Bottom Panel: $<H_{SHO}^s>$ (dashed), $<H_e>$ (dotted) and $<H_w> \equiv <H_s> + <H_e> + <H^I>$ (solid).  All the quantities show the expected physical behavior until $t \approx 10^{14}$ where the breakdown of the numerical computation of the phases sets in. This figure illustrates the very large dynamic range of our numerical computations. (Recall that the SHO period is $2\pi$.) }
    \label{fig:NumFail}
\end{figure}

Regarding numerical accuracy, the critical aspect was the ability of our code to accurately evaluate exponentials with potentially large imaginary arguments (to rotate the phases). The residuals shown in Figs.~\ref{fig:SHOcstest} and~\ref{fig:SHOsc} give some sense of the capabilities of our code. Note that while those figures refer to the case where $E_I=0$ and focus on the behavior of the SHO, the results were generated with $E^e=0.03$ and $N_e=600$ (and thus $N_w=18,000$) so the residuals reflect a stronger test than one might initially expect. Figure~\ref{fig:NumFail} shows several quantities discussed in this paper evolved to later times than previously shown. One can see evidence of the breakdown of numerical accuracy around $t=10^{14}$, when the exponential expressions for the (extremely large) phases start failing to compute properly.  For example energy conservation (the constancy of the solid curve in the lower panel) is lost, and the requirement that $S \leq S_{max} =  \ln\left(N_s\right)$ (Eqn.~\ref{eqn:Smaxdef}) is violated.  These, and many other tests of the numerics proved robust up to times just below the $t\approx10^{14}$ breakdown point. The availability of accurate numerical computations over such a wide time range provides excellent latitude for exploring the physics of the ACL model. (For context, recall that the period of the oscillator is $\tau = 2\pi$.)\footnote{
For the senior member of this collaboration whose last experience with this kind of calculation was in the 1990's~\cite{Albrecht:1992rs} the comparison of capabilities between then and now is truly remarkable.}

Our calculations were performed using Matlab on a 64 bit Windows computer with a 3.6GHz Intel i7-4790 processor and 32GB RAM. Each time step, which included calculating a wide variety of information from $\rho_s$ and $\rho_e$ (including the sort reported here), took 20-30 seconds. (We noticed a roughly 25\% speedup after simultaneously upgrading from Windows 8.1 to 10 and from Matlab R17a to R18b.) The initial construction of all relevant matrices (of which the diagonalization of $H_w$ is the most time consuming) takes around $1.5$ hours for the case with $N_s=30$ and $N_e = 600$.  We rarely wanted more than 2000 time steps to produce long times views such as shown in Figs.~\ref{fig:EQMbE} and~\ref{fig:EQMbES}, and for many purposes (such as Fig.~\ref{fig:Moments} and various rough explorations) a lot fewer were sufficient. Much of our code development and testing could be done with smaller environment sizes, for which the time steps were more or less instantaneous.  With these sorts of turnaround times we found it possible to work with the ACL model in a reasonably interactive manner.

\clearpage
\bibliography{AAlib}
\end{document}